\documentclass[12pt]{article}
\usepackage[super,numbers,sort&compress]{natbib}
\usepackage{times}
\usepackage[version=3]{mhchem} 
\usepackage[]{amsmath} 
\usepackage{physics}
\usepackage{graphicx}
\usepackage{caption}
\title
{Radiative properties of quantum emitters in boron nitride from excited state calculations and Bayesian analysis}

\author{Shiyuan Gao,$^1$ Hsiao-Yi Chen,$^{1,2}$ Marco Bernardi$^{1,\ast}$\\
\\
\normalsize\emph{$^1$ Department of Applied Physics and Material Science, California Institute of Technology,}\\
\normalsize\emph{Pasadena, California 91125, USA}\\
\normalsize\emph{$^2$ Department of Physics, California Institute of Technology,}\\
\normalsize\emph{Pasadena, California 91125, USA}\\
\\
\normalsize{$^\ast$To whom correspondence should be addressed; E-mail:  bmarco@caltech.edu.}
}

\date{}

\begin{document}


\baselineskip24pt


\maketitle 
\begin{abstract}
\baselineskip20pt
Point defects in hexagonal boron nitride (hBN) have attracted growing attention as bright single-photon emitters. 
However, understanding of their atomic structure and radiative properties remains incomplete. 
Here we study the excited states and radiative lifetimes of over 20 native defects and carbon or oxygen impurities in hBN using ab initio density functional theory and GW plus Bethe-Salpeter equation calculations, 
generating a large data set of their emission energy, polarization and lifetime. 
We find a wide variability across quantum emitters, with exciton energies ranging from 0.3 to 4 eV and radiative lifetimes from ns to ms for different defect structures. 
Through a Bayesian statistical analysis, we identify various high-likelihood defect emitters, among which the native $\mathrm{V_NN_B}$ defect is predicted to possess emission energy 
and radiative lifetime in agreement with experiments. 
Our work advances the microscopic understanding of hBN single-photon emitters and introduces a computational framework to characterize and identify quantum emitters in 2D materials.
\end{abstract}
%

\makeatletter
\let\origsection\section
\renewcommand\section{\@ifstar{\starsection}{\nostarsection}}

\newcommand\nostarsection[1]
{\sectionprelude\origsection{#1}\sectionpostlude}

\newcommand\starsection[1]
{\sectionprelude\origsection*{#1}\sectionpostlude}

\newcommand\sectionprelude{%
 \vspace{8pt}
}

\newcommand\sectionpostlude{%
  \vspace{-8pt}
}
\makeatother

\section*{Introduction}

Two-dimensional (2D) materials are rapidly becoming a new platform for photon based quantum information technologies \cite{Liu2019,XChen2019}. 
The discovery of single-photon emitters (SPEs) at point defects in hBN \cite{Tran2016} has spurred an intense search for optically active defects in 2D crystals \cite{Gottscholl2020}. 
Compared to defects in bulk crystals, such as diamond and silicon carbide \cite{Childress2006, Ladd2010, Hensen2015, Aharonovich2016},
defects embedded in 2D materials promise to be more easily addressed and controlled.   
In the case of hBN, defect emitters exhibit a range of desirable properties, including high emission rate, room temperature stability, strong zero-phonon line (ZPL) and easy integration with other optical components using 2D hBN crystals \cite{Liu2019,Caldwell2019,Jungwirth2016,Tran2017,Dietrich2018}. 
\\ 
\indent 
A pressing challenge for defect emitters in hBN is identifying their atomic structure. Various possible structures have been proposed on the basis of density functional theory (DFT) calculations 
and their comparison with experiments \cite{Tran2016,Tawfik2017,Weston2018,Sajid2018,Lopez-Morales2018,Noh2018,Abdi2018,Turiansky2019}. 
However, while DFT can provide valuable insight into the formation energy, symmetry and electronic structure of SPE defects, 
it cannot address key aspects of point-defect SPEs such as their excited states and radiative processes responsible for light emission.
To complicate the matter further, similar to other 2D materials and their defects \cite{Bernardi2013, Qiu2013, Refaely-Abramson2018}, 
optical transitions at defects in hBN are dominated by excitonic effects \cite{Attaccalite2011}, which require specialized first-principles calculations beyond the scope of DFT. Quantum chemistry approaches and density matrix renormalization group have also been used to investigate specific defect structures \cite{Reimers2018,Ivady2020}, but wide comparisons among different structures are still missing.
\\
\indent
The Bethe-Salpeter equation (BSE) can accurately predict optical properties and excitons in materials \cite{Rohlfing}. 
It also enables, due to recent advances, precise calculations of radiative lifetimes in 2D and bulk crystals \cite{Palummo2015, Chen2018, Chen2019, Jhalani2019}. 
The radiative lifetime plays an important role in the study of SPEs as it determines the shortest decay time constant in the second-order photon correlation function \cite{Loudon, Kimble} and can also be measured directly from fluorescence intensity decay \cite{Tran2016, Jungwirth2016}. 
Applying the BSE and related methods to defect emitters in hBN would enable direct comparisons between theory and experiment of the emission energy and radiative lifetime,  providing valuable information to identify defect SPEs. 
Yet, first-principles BSE calculations are computationally costly to carry out on defect structures, 
and the radiative lifetime calculations are only a recent development. 
\\ 
\indent
In this work, we employ the BSE approach to compute from first principles the optical properties, transition dipoles, excitons and radiative lifetimes of atomic defects in hBN. 
We examine a large pool of candidate SPE structures, spanning native defects and carbon or oxygen impurities, to correlate their atomic structures with their photophysics. 
We find that different quantum emitters exhibit radiative lifetimes spanning six orders of magnitude and emission energies from infrared to ultraviolet. 
Bayesian statistical analysis is employed to correlate our results with experiments and identify the most likely SPEs in hBN, among which we find the $\mathrm{V_NN_B}$ defect to have the highest likelihood.  
In-depth calculations on the $\mathrm{V_NN_B}$ defect highlight the strong dependence of its radiative properties on small perturbations to its atomic structure. 
The dependence of the defect radiative properties on dielectric screening is analyzed by comparing monolayer and bulk hBN results.
Our systematic investigation addresses key challenges for characterizing the excited states and radiative properties of defect emitters in 2D materials.  

\section*{Results}

\begin{figure}[t!]
\centering
\includegraphics[width=1.0\textwidth,clip]{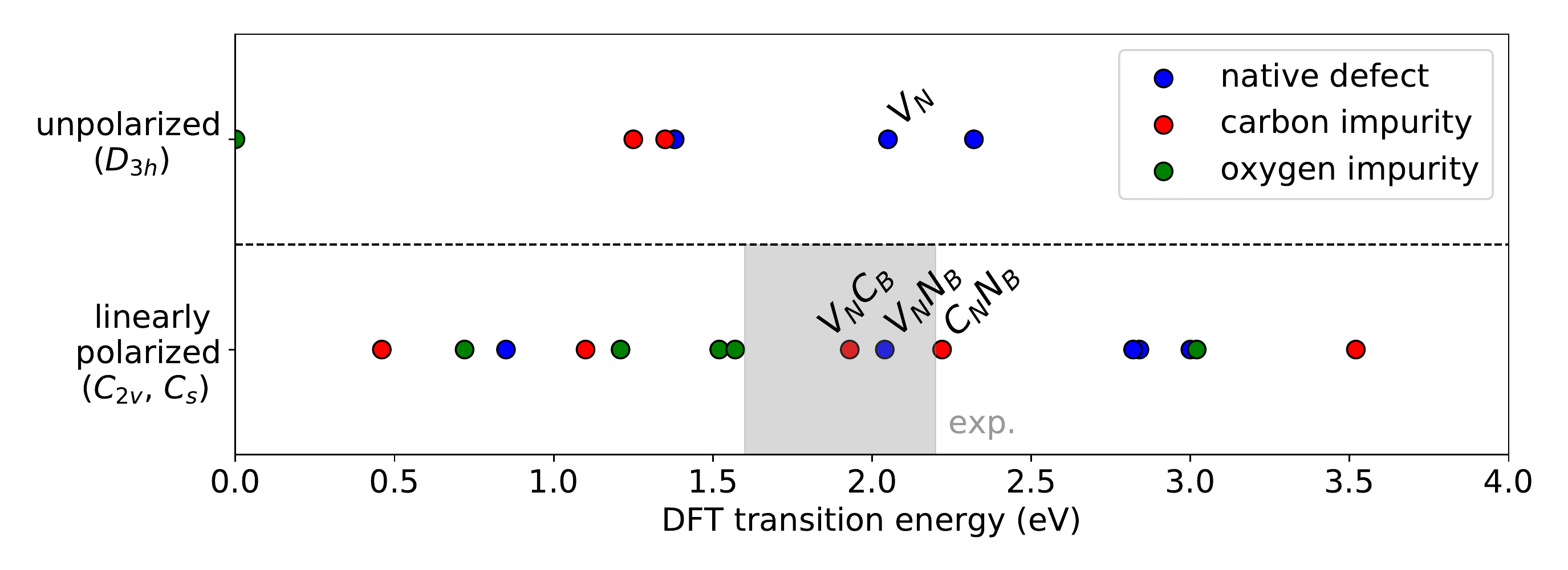}
\caption{Figure 1. Distribution of the DFT transition energy, structural symmetry and emitted light polarization of 22 candidate defect structures. The shaded area shows the experimental range of values for SPEs in hBN.}
\label{fig:1}
\end{figure}

Our candidate defect structures consist of charge-neutral native defects and carbon or oxygen impurities occupying one or two atomic sites, for a total of 8 different native defects and 7 structures for each of carbon and oxygen impurities. 
We compute the ground state defect properties using DFT, employing fully relaxed defect atomic structures in 5$\times$5 supercells of monolayer hBN (and for some defects, in bulk hBN).  
We then refine the electronic structure of selected defects using GW calculations \cite{Hybertsen1986}, followed by BSE calculations to obtain the exciton energies and wave functions and from them the optical absorption, transition dipoles and radiative lifetimes (see the Methods section). 
In the following, we denote the defects in hBN as $\mathrm{X_NY_B}$ if neighboring N and B atoms are replaced by species X and Y, respectively, where X and Y can be a vacancy or another element \cite{Tawfik2017,Tran2016}. 
We focus on emitters in the interior of the 2D crystal \cite{Exarhos2017,Hayee2020} and do not consider defects that would likely appear at the sample edges or corners \cite{Chejanovsky2016,Choi2016}. 
\\
\indent
The electronic energies obtained using DFT, while in general not representative of electronic or optical transitions, can be used for guidance and for estimating qualitative trends.  
Figure \ref{fig:1} shows the lowest spin-conserving transition (HOMO-LUMO) energy of the candidate defects, obtained from DFT,     
together with the emission polarization inferred from structural symmetry. The defect structures considered here exhibit three different types of local symmetries, $\mathrm{D_{3h}}$, $\mathrm{C_{2v}}$, and $\mathrm{C_s}$. 
In the high symmetry $\mathrm{D_{3h}}$ structure, adopted by $\mathrm{N_B}$, $\mathrm{V_N}$, $\mathrm{B_N}$, $\mathrm{C_B}$, $\mathrm{C_N}$, $\mathrm{O_N}$, emitted light cannot be linearly polarized. 
Conversely, linearly polarized emitted light, as observed experimentally in hBN SPEs \cite{Tran2017}, is possible in the $\mathrm{C_{2v}}$ and $\mathrm{C_s}$ symmetries. 
In the $\mathrm{C_{2v}}$ configuration, which is the most common among the defects investigated here, the 3-fold rotational symmetry is broken but all the atoms remain in-plane, preserving the mirror symmetry with respect to the crystal plane. 
The $\mathrm{C_s}$ symmetry found in the $\mathrm{V_NN_B}$, $\mathrm{V_NC_B}$, $\mathrm{V_NO_B}$ and $\mathrm{O_B}$ defects is instead associated with an out-of-plane distortion that breaks the mirror symmetry about the plane. 
The DFT transition energies for the 22 defect structures range from 0 to 3.5 eV. In contrast, the ZPL of the measured SPEs are in the 1.6$-$2.2 eV energy range \cite{Tran2017}, as shown by the shaded region in Fig.~\ref{fig:1}. 
While candidate structures with $\mathrm{D_{3h}}$ symmetry can be ruled out, $\mathrm{C_{2v}}$ and $\mathrm{C_s}$ structures with exceedingly small or large DFT transition energies also appear unlikely on the basis of the DFT results. 

Starting from the DFT ground state, for selected defects we compute the excited state properties with the GW-BSE method, obtaining the quasiparticle energies in the one-shot $\mathrm{G_0W_0}$ approximation 
and the exciton energies and wave functions with the BSE, which captures electron-hole interaction and excitonic effects. 
We combine the solutions of the BSE with an approach we recently developed \cite{Chen2019,Jhalani2019} to compute the radiative lifetime of an exciton state from Fermi's golden rule. 
Generalizing our previous formula for isolated (0D) emitters \cite{Chen2019} to include anisotropic dielectric screening in hBN, the radiative decay rate $\gamma_S$ (inverse of radiative lifetime) of an exciton state $S$ is:
\begin{equation}
  \gamma_S=\frac{\sqrt{\epsilon_{xy}(k_{xy})}e^2E_S}{3\pi\epsilon_0m^2c^3\hbar^2}\left[\left(\frac{3}{4}+\frac{\epsilon_z}{4\epsilon_{xy}(k_{xy})}\right)|p_{S,xy}|^2+|p_{S,z}|^2\right]\,,
\label{eq:lifetime}
\end{equation}
where $\epsilon_{xy}(k_{xy})$ and $\epsilon_{z}$ are the in-plane and out-of-plane dielectric function of hBN, respectively, $k_{xy}$ is the in-plane photon wavevector, $E_S$ is the exciton energy and $p_{S,xy}$ and $p_{S,z}$ are the corresponding components of the exciton transition dipole. 
We use a constant in-plane dielectric function for bulk hBN ($\epsilon_{xy} = 5$) at optical wavelengths, and for monolayer hBN we take into account the dependence on wavevector $q$ as 
$\epsilon_{xy}(q)\approx 1+2\pi\alpha_{2D}\,q$ \cite{Cudazzo2011}, where $\alpha_{2D}$ is a constant equal to 0.4 nm \cite{Andersen2015}. 
In this approach, which is appropriate for 2D materials, the in-plane dielectric function of monolayer hBN reduces to a value of 1 when the wavevector $q$ equals the wavevector of a photon at optical frequencies.
\\
\indent 
\begin{figure}[t!]
\centering
\includegraphics[width=1.0\textwidth,clip]{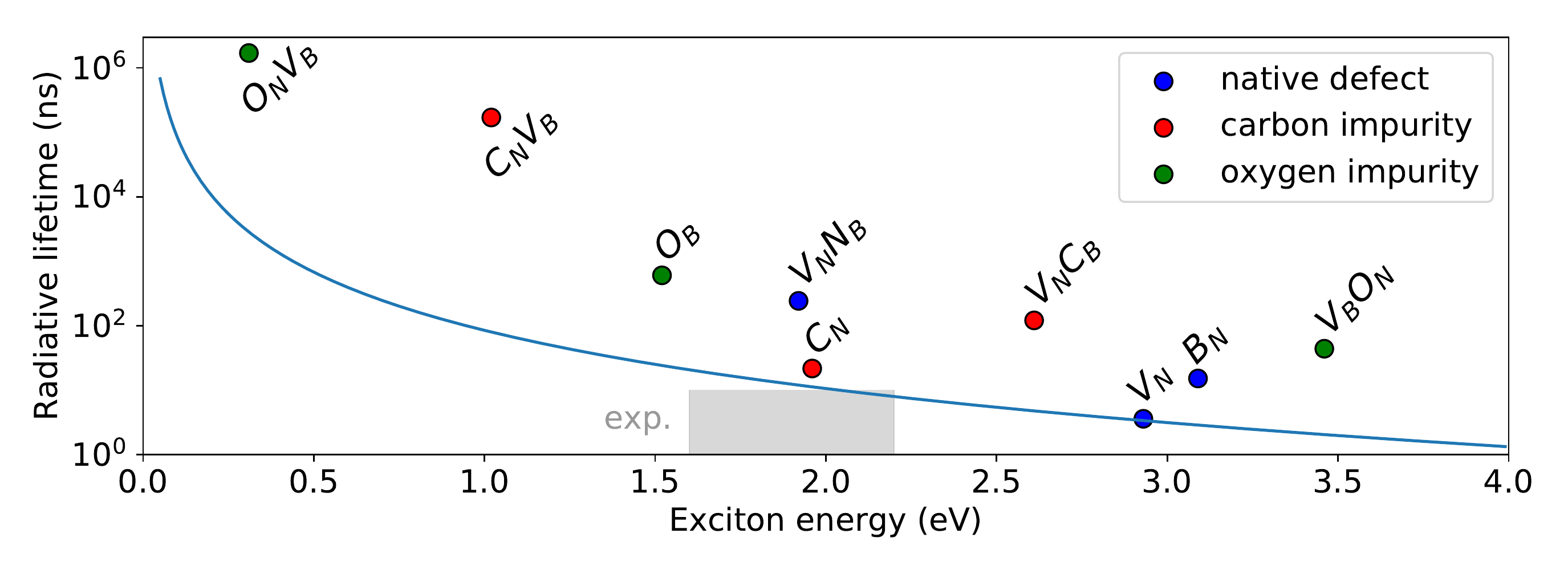}
\caption{Figure 2. Radiative lifetime and energy of the lowest bright exciton of candidate defect SPEs in hBN from GW-BSE calculations. The range of experimental values is shown as a shaded region. The blue line shows the radiative lifetime of an exciton with an assumed transition dipole moment of 6.9 Debye.}
\label{fig:2}
\end{figure}
Figure \ref{fig:2} shows the computed radiative lifetimes and lowest bright exciton energies for nine selected defects, including $\mathrm{V_N}$, $\mathrm{B_N}$, $\mathrm{V_NN_B}$, $\mathrm{C_N}$, $\mathrm{V_NC_B}$, $\mathrm{C_NV_B}$, 
$\mathrm{O_B}$, $\mathrm{O_NV_B}$ and $\mathrm{V_BO_N}$. We find that the exciton energy can differ significantly $-$ by as much as 1 eV $-$ from the corresponding DFT transition energy, which fails to account for screening and electron-hole interaction effects. 
This result is a testament to the importance of excited state calculations for investigating SPEs.\\
\indent  
Surprisingly, we find that the computed radiative lifetime for the selected structures span six orders of magnitude, from about 1 to 10$^6$ ns (1 ms), 
showing that the emission rate and brightness of quantum emitters in hBN can vary widely. 
The values typically found in experiments for the emission energy (1.6$-$2.2 eV) and radiative lifetime (1$-$10 ns) are also given for comparison in Fig.~\ref{fig:2}. 
Writing the exciton transition dipole in Eq. 1 as $\mathbf{p}_S = -(i m E_S / \hbar e) \times e\, \mathbf{r}_S$ to highlight its physical meaning of an atomic-scale dipole, 
and setting $\abs{\mathbf{r}_S} = \abs{\bra{0}\mathbf{r}\ket{S}}$ equal to the in-plane B-N bond length (this choice gives a dipole of 6.9 Debye), the resulting radiative lifetime as a function of exciton energy gives a lower bound to the calculated radiative lifetimes (see the blue line in Fig.~\ref{fig:2}). The physical insight of this analysis is that due to incomplete overlap of the electron and hole wavefunctions, the exciton transition dipole for most defects is significantly smaller than the bond length, leading to longer radiative lifetimes than this theoretical bound. 
\\
\indent
As the candidate defect structures have properties distributed across a wide range, it is challenging to pinpoint the correct defect structure from our results.  
As such, we pursue a quantitative analysis of the relative likelihood of the various structures using Bayesian inference, 
a statistical approach for dealing with uncertainties and combining information from different categories, in which the probability of a hypothesis is updated as more evidence or information becomes available \cite{VonToussaint2011}.
The workflow of our \mbox{analysis is shown in Fig.~\ref{fig:3}.}
%
%
We first generate candidate structures systematically by enumerating all defects that occupy no more than two adjacent atomic sites. 
A prior likelihood $p(h)$ is then assigned to each candidate structure $h$ as an initial guess of how likely the defect structure can appear in the hBN crystal without considering whether it could account for the properties of the SPEs. The prior likelihood can be tailored to describe various experimental scenarios. For example, if carbon impurities are added in the experiment, as shown in recent work \cite{Mendelson2020}, one could include the presence of a carbon impurity in the prior likelihood, therefore favoring the posterior likelihood of carbon-containing emitters. 
Here, without referring to specific experimental conditions, we choose the prior likelihood $p(h)$ on the basis of the structural complexity of the defect, using the single exponential form $p(h)\propto S_hA^{-C_h}$, 
where $C_h$ measures the complexity of the structure, defined as the number of vacancies and antisite defects plus twice the number of impurities,  
$A$ is a constant, and $S_h$ is a degeneracy factor due to symmetry-related configurations. 

We subsequently combine our computational results with experimental evidence to obtain a posterior likelihood $p(h|E)$ for each structure \cite{VonToussaint2011}:
\begin{equation}
  p(h|E)\propto p(h)p(E|h),
\label{eq:bayesian}
\end{equation}
where the likelihood function $p(E|h)$ is the probability that the structure $h$ is compatible with a specific piece of experimental evidence $E$ on the basis of our calculations. 
The resulting posterior likelihood, which is updated in multiple steps using several computed properties, quantifies which structures among those considered are more likely to be defect emitters in hBN and deserve further consideration. 

\begin{figure}[h!]
\centering
\includegraphics[width=1.0\textwidth,trim={1cm 2.5cm 2.5cm 1.5cm},clip]{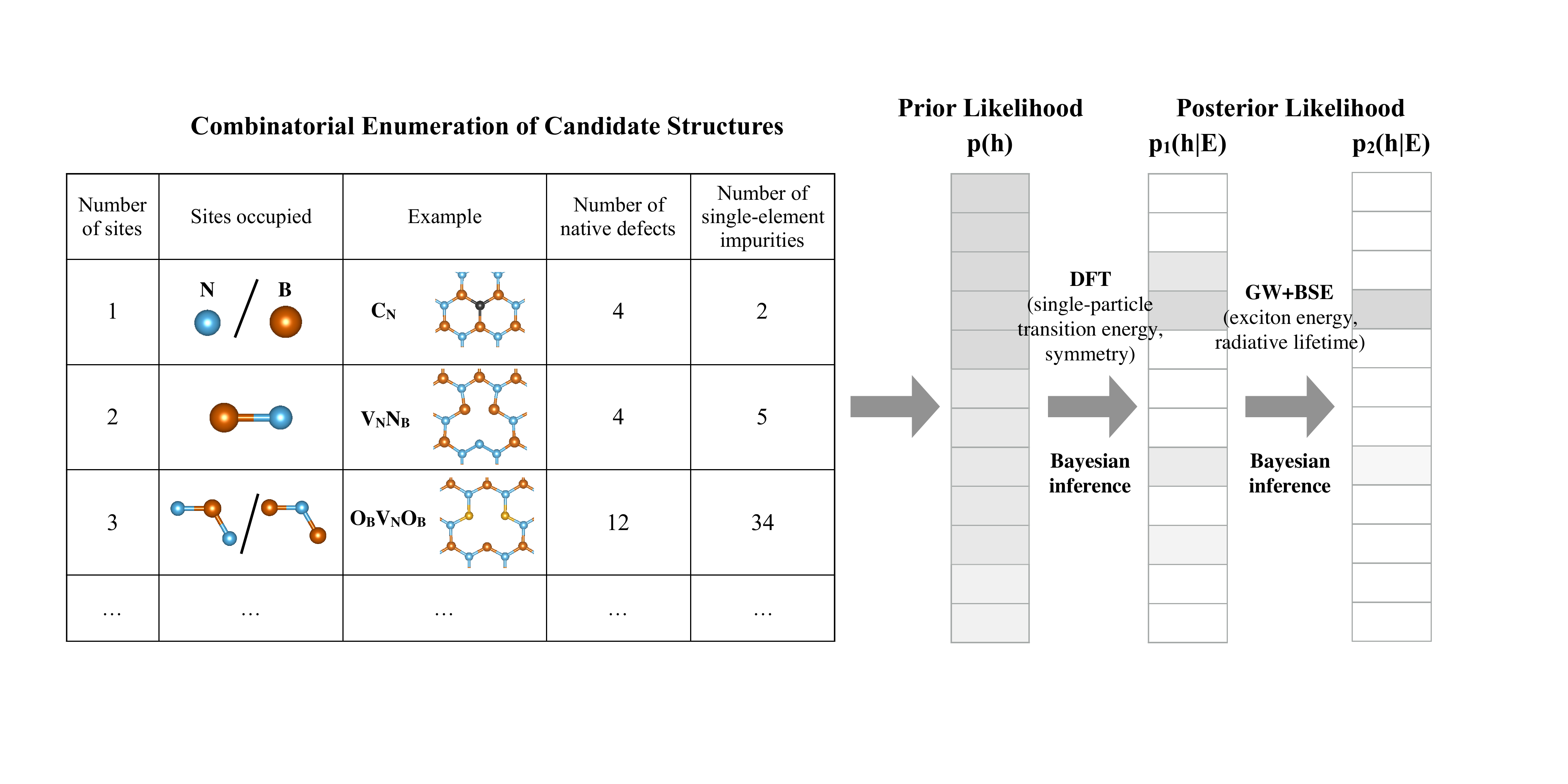} 
\caption{Figure 3. Bayesian inference workflow. Candidate structures are initially generated through combinatorial enumeration of point defects in hBN and attributed a prior likelihood based on structural complexity. 
The likelihood of each defect is then updated using results from our DFT calculations. For the most promising defect structures, the likelihood is refined using GW-BSE calculations.}
\label{fig:3}
\end{figure}

The properties of the candidate SPE defects and the results of our Bayesian inference analysis are summarized in Table 1. 
Moving in the table from left to right, the prior probability $p(h)$ is updated to posterior probabilities $p(h|E)$ using calculations presented in subsequent columns. 
The first computed posterior probability, $p_1(h|E)$, uses the DFT results to infer whether the defect symmetry is compatible with linearly polarized emission   
and whether the DFT transition energy, within a standard deviation of $\sim$0.5 eV, falls in the 1.6$-$2.2 eV emission range found in experiments \cite{Jungwirth2016,Exarhos2017,Tran2017}. 
The final posterior likelihood $p_2(h|E)$ in the rightmost column of Table 1 gives the likelihood of the most likely defect structures when all factors have been taken into account, 
including comparison between the lowest bright exciton energy and the experimental emission energy and comparison with experiment of the radiative lifetime from our GW-BSE calculations. 
The detailed rationale of our Bayesian analysis is described in the Supplementary information.
\\
\indent
The main finding from the data in Table 1 is that the $\mathrm{V_NN_B}$ defect, which was originally proposed as a SPE in hBN on the basis of DFT calculations \cite{Tran2016,Tawfik2017,Abdi2018}, 
possesses optical and radiative properties that most accurately match the experimental results, even after taking into account excitonic effects and radiative lifetimes. 
The next most likely structure is the oxygen impurity defect, $\mathrm{O_B}$, with an emission energy of 1.52 eV just below the experimental range. 
On the other hand, we find that the $\mathrm{V_NC_B}$ defect, which is also considered a likely candidate in the literature based on DFT calculations \cite{Tawfik2017,Abdi2018}, 
has an emission energy of 2.6 eV that lies too far above the experimental energy range when excitonic effects are included, making it a less likely candidate. 
We note that although our analysis excludes the $\mathrm{V_NC_B}$ defect and other defects with higher emission energies as candidates for the 1.6$-$2.2 eV SPEs, they could still be good candidates for SPEs in the ultraviolet range\cite{bourrellier2016,tan2019}, which is not the focus of our discussion. 
As the most likely defect emitter in hBN is the $\mathrm{V_NN_B}$ structure according to our analysis, 
we investigate it in more detail to provide a microscopic understanding of its radiative properties.  

\begin{figure}[t!]
\centering
\includegraphics[width=1.0\textwidth,trim={0cm 3.0cm 0cm 2cm},clip]{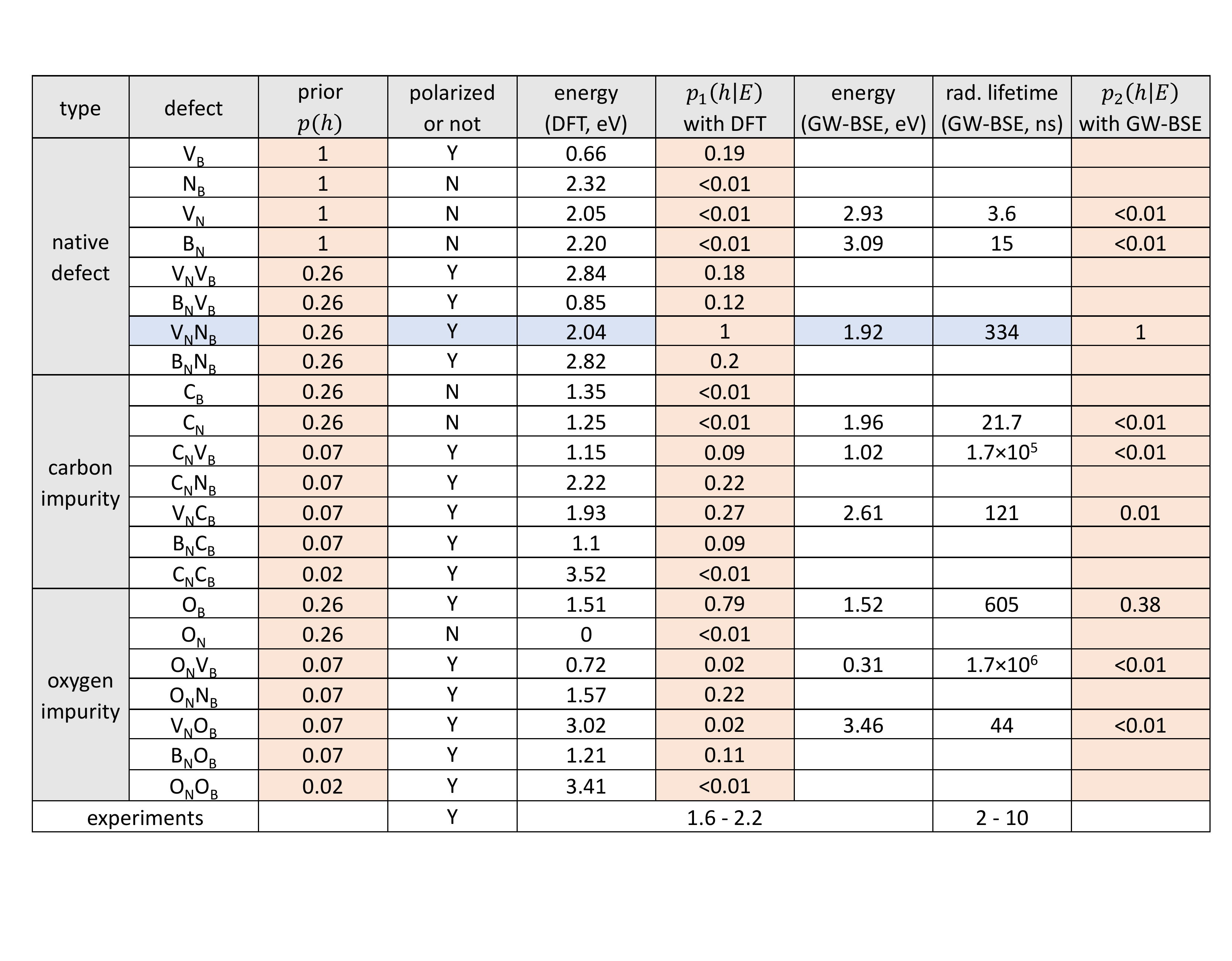}
\caption[]{Table 1. Calculated properties of candidate defect emitters in hBN and their likelihood based on Bayesian inference analysis. The likelihood is normalized to a maximum \mbox{value of 1.}}
\label{table:1}
\end{figure}

Figure 4(a) shows that the relaxed atomic structure of the $\mathrm{V_NN_B}$ defect in monolayer hBN exhibits an out-of-plane displacement of the central nitrogen (N) atom by $z=0.66$~\AA. 
If the structure is made planar by constraining the N atom to the hBN plane, the total energy of the system is only moderately higher (by 0.125 eV) than the equilibrium structure.  
Therefore the total energy forms a shallow double-well potential as a function of the out-of-plane N atom displacement. 
Consistent with recent findings \cite{Li2020}, this small energy difference suggests that the structure is soft in the out-of-plane direction and could fluctuate due to external perturbations. 
\\
\indent
To take this result into account, we compute the optical absorption spectrum of the $\mathrm{V_NN_B}$ defect for out-of-plane displacements $z$ of the N atom in the 0$-$0.83~\AA~range, 
interpolating between the in-plane and out-of-plane equilibrium structures to obtain the intermediate metastable structures. 
We find that the absorption spectrum changes drastically for even small changes of $z$, as shown in Fig.~4(c). The spectrum is dominated by two exciton absorption peaks at low energy. 
At the equilibrium position ($z=0.66$~\AA), the first peak at 1.92 eV is associated with a transition from the top valence to the bottom conduction spin-majority electronic states, whose wave functions are shown in Fig.~4(b). 
As the displacement $z$ of the N atom is decreased from the equilibrium value to zero, which corresponds to a planar structure, the energy of the first exciton peak decreases at first 
but then increases again at small $z$ values while gaining oscillator strength monotonically.
The second peak at 2.7 eV is also due to a transition from the top valence to the bottom conduction spin-minority bands with relatively large oscillator strength at the out-of-plane equilibrium N atom position. 
Both the energy and the oscillator strength of this peak decrease monotonically as $z$ decreases [see Fig.~4(c)]. 
For the planar structure ($z$=0), the second peak becomes the lowest-energy transition but completely loses its oscillator strength and becomes a dark state. 

\begin{figure}[h!]
\centering
\includegraphics[width=1.0\textwidth,trim={3cm 0cm 5cm 1cm},clip]{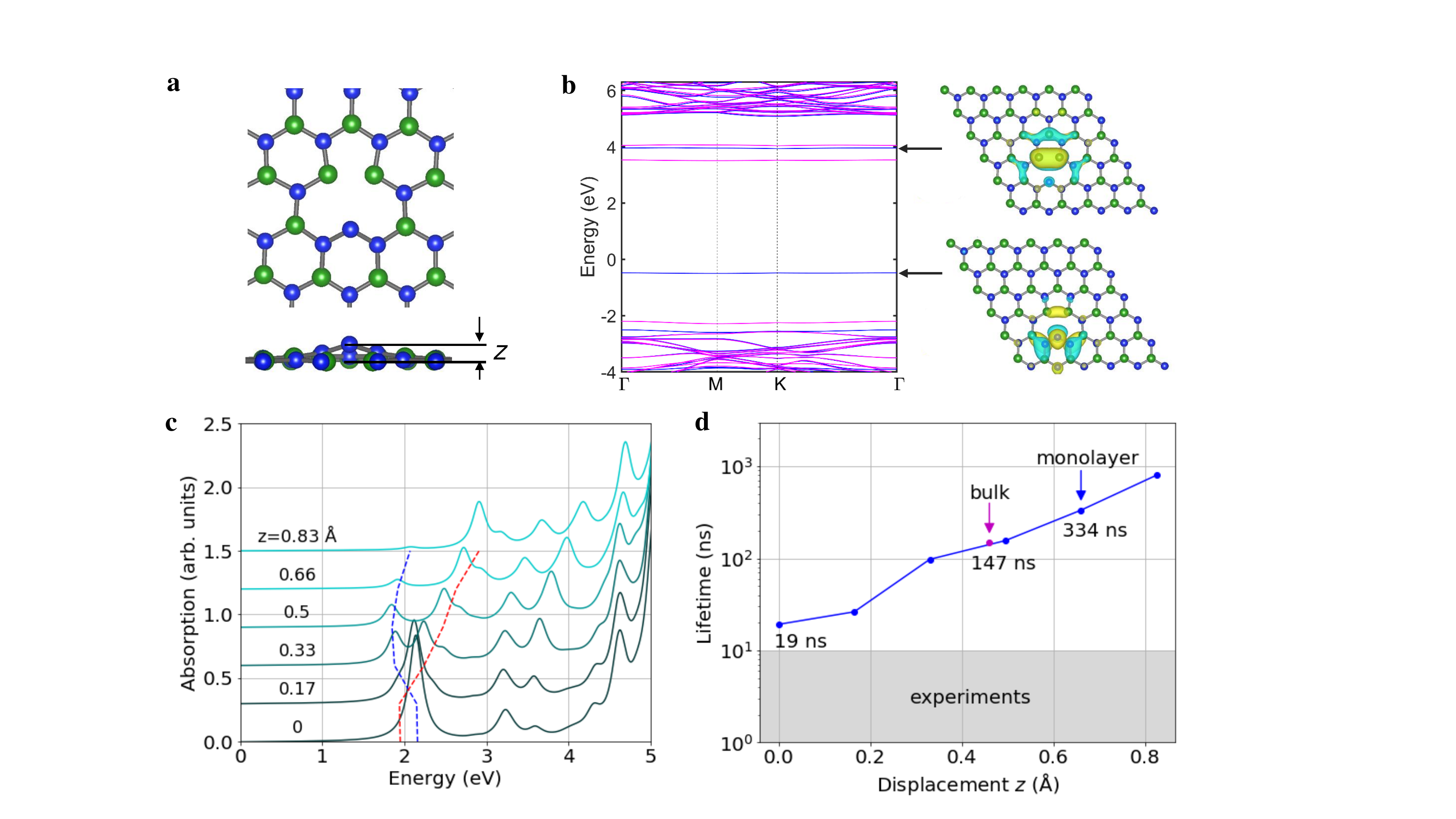}
\caption{Figure 4. Properties of the $\mathrm{V_NN_B}$ defect. (a) Atomic structure in top and side views, with N atoms shown as blue and B atoms as green spheres. 
Also shown in the side view is the N atom out-of-plane displacement $z$. (b) GW electronic band structure computed using a $5\times5$ supercell. 
The wave function of the two majority-spin bands that mainly contribute to the lowest-energy exciton are plotted on the right. 
(c) Absorption spectrum and (d) radiative lifetime as a function of the out-of-plane displacement $z$ of the central N atom.}
\label{fig:4}
\end{figure}

These noteworthy changes in the two lowest-energy exciton peaks are accompanied by substantial changes in the exciton radiative lifetimes. 
Although the lifetime of the lowest exciton is 334 ns for the DFT equilibrium out-of-plane distortion, it becomes only 19 ns if the $\mathrm{V_NN_B}$ structure is kept planar ($z=0$ case), as shown in Fig.~4(d).  
This value of the radiative lifetime for the planar structure is within a factor of 2$-$3 of experimental results \cite{Tran2017}. 
Because many experiments are carried out on thicker hBN samples, we also calculated the radiative lifetime of the $\mathrm{V_NN_B}$ defect embedded in bulk hBN (instead of monolayer), 
a setting resembling more closely SPE experimental measurements in thicker hBN layer stacks (see the Methods section). Due to interlayer interactions, the out-of-plane displacement of the N atom is reduced to $0.45$~\AA~in bulk hBN. 
Going from monolayer to bulk reduces the radiative lifetime of the $\mathrm{V_NN_B}$ defect to 147 ns due to changes in the dielectric environment and transition dipole, both of which affect the radiative emission rate in Eq.~(1).
Therefore we conclude that the radiative lifetime of the $\mathrm{V_NN_B}$ defect is highly sensitive to both out-of-plane structural distortions and the dielectric environment.

\section*{Discussion}

While our computed lifetimes are in the 20$-$350 ns range for different $\mathrm{V_NN_B}$ structures, these values pertain to an ideal defect with 100\% quantum yield, 
whereas defect emission measured so far in hBN is likely associated with a lower quantum yield. To explain the discrepancy between the experimental and calculated lifetimes, note that nonradiative processes, 
which are not included in our calculation of intrinsic radiative lifetimes, could significantly reduce the apparent lifetime measured in experiments. 
For example, a defect with a measured 10 ns lifetime but only a 10\% quantum yield would correspond to a 100 ns intrinsic radiative lifetime in very good agreement with our calculations. 
Future SPE measurements in hBN taking into account the quantum yield may provide a better estimate of the intrinsic SPE radiative lifetime and enable a more accurate comparison with our calculations. 
\\
\indent 
In summary, we have investigated the excited state and radiative properties of many candidate defect SPEs in hBN. 
Our calculations address the photophysics of these defect structures, including their optical transitions and radiative lifetimes, 
using calculations that accurately account for excitonic effects and anisotropic dielectric screening. 
Our Bayesian statistical analysis allows us to identify the native $\mathrm{V_NN_B}$ defect as the most likely candidate within a wide pool of over 20 charge-neutral native defects and carbon or oxygen impurities. 
This work explores the fertile ground at the intersection of excited-state first-principles calculations and Bayesian learning methods, and the application of this framework to quantum technologies.    
The search for novel single emitters will greatly benefit from such precise first-principles calculations of excited states in materials.

\section*{Methods}

We carry out DFT calculations in the generalized gradient approximation using the Perdew-Burke-Ernzerhof (PBE) exchange-correlation functional \cite{Perdew1996}. 
Our spin-polarized DFT calculations employ the plane-wave pseudopotential method implemented in {\sc Quantum Espresso} \cite{QE}. 
The defects are placed in a 5$\times$5 supercell of hBN with the lattice constant kept fixed at 5$\times$2.504~\AA ~while the atoms are fully relaxed without symmetry constraints. 
For calculations on monolayer hBN, a vacuum of 15~\AA ~is used to avoid spurious inter-layer interactions with the periodic replicas.  
We use ONCV pseudopotentials \cite{Hamann2013,Schlipf2015} for all atoms along with a plane-wave kinetic energy cutoff of 80 Ry and a 3$\times$3 $\mathbf{k}$-point Brillouin zone grid. 
The GW-BSE calculations are carried out with the Yambo code \cite{Marini2009,Sangalli2019} using a 2D slab cutoff of the Coulomb interaction. 
For calculations of defects in monolayer hBN, a 3$\times$3 $\mathbf{k}$-point grid is employed together with an energy cutoff of 10 Ry for the dielectric matrix. 
The number of empty bands included in the GW calculation (for the polarizability and self-energy summations) is 7 times the number of occupied bands. 
These parameters converge the exciton energy to within 0.1~eV in our test calculations, which are shown in the Supplementary information. 
For calculations on defects in bulk hBN, we use a 5$\times$5$\times$1 supercell containing two AA’-stacked layers. 
The bulk case uses a 2$\times$2$\times$3 $\mathbf{k}$-point grid; all other parameters are the same as in the monolayer case.

\section*{Data Availability} 

All data needed to evaluate the conclusions in the paper are present in the paper and/or the Supplementary information. Additional data related to this paper may be requested from the authors.

\section*{Acknowledgements} 
We acknowledge valuable discussions with Hamidreza Akbari. \\
This work was supported by the Department of Energy under Grant No. DE-SC0019166. The radiative lifetime code development was partially supported by the National Science Foundation under Grant No. ACI-1642443. 
H.-Y.C. was partially supported by the J. Yang Fellowship. This research used resources of the National Energy Research Scientific Computing Center, 
a DOE Office of Science User Facility supported by the Office of Science of the US Department of Energy under Contract No. DE-AC02-05CH11231. 

\section*{Author contributions} 
S.G. and M.B. conceived and designed the research. S.G. performed calculation and analysis. H.-Y.C. provided theoretical support. M.B. supervised the entire research project. All authors discussed the results and contributed to the manuscript. 

\section*{Competing interests} 
The authors declare that they have no competing financial interests.

\section*{Additional information} 
Supplementary information is available for this paper (at URL to be added).

\end{document}